\documentclass[9pt,twocolumn,twoside]{opticajnl}
\journal{opticajournal} % use for journal or Optica Open submissions

\setboolean{shortarticle}{true}
\usepackage{siunitx}
\usepackage{xurl}

\usepackage{lineno}
\title{Single-shot X-ray ptychography as a structured illumination method}

\author[1,6]{Abraham Levitan}
\author[1]{Klaus Wakonig}
\author[1,2]{Zirui Gao}
\author[1,3]{Adam Kubec}
\author[4]{Bing Kuan Chen}
\author[4]{Oren Cohen}
\author[1,5,*]{Manuel Guizar-Sicairos}

\affil[1]{Paul Scherrer Institute, Forschungsstrasse 111, 5232 Villigen PSI,Switzerland}
\affil[2]{Brookhaven National Laboratory, Upton, NY 11973- 5000, USA}
\affil[3]{Current address: XRnanotech AG, Parkstrasse 1, 5234 Villigen, Switzerland}
\affil[4]{Solid State Institute and Physics Department, Technion-Israel Institute of Technology, Haifa, 3200003, Israel}
\affil[5]{Institute of Physics, École Polytechnique Fédérale de Lausanne (EPFL), Lausanne, Switzerland}
\affil[6]{abraham.levitan@psi.ch}
\affil[*]{manuel.guizar-sicairos@psi.ch}

\begin{abstract}
Single-shot ptychography is a quantitative phase imaging method wherein overlapping beams of light arranged in a grid pattern simultaneously illuminate a sample, allowing a full ptychographic dataset to be collected in a single shot. It is primarily used at optical wavelengths, but there is interest in using it for X-ray imaging. However, the constraints imposed by X-ray optics have limited the resolution achievable to date. In this work, we reinterpret single-shot ptychography as a structured illumination method by viewing the grid of beams as a single, highly structured illumination function. Pre-calibrating this illumination and reconstructing single-shot data using the randomized probe imaging algorithm allows us to account for the overlap and coherent interference between the diffraction arising from each beam. We achieve a resolution 3.5 times finer than the numerical aperture-based limit imposed by traditional algorithms for single-shot ptychography. We argue that this reconstruction method will work better for most single-shot ptychography experiments and discuss the implications for the design of future single-shot X-ray microscopes.
\end{abstract}

\setboolean{displaycopyright}{false} % Do not include copyright or licensing information in submission.

\begin{document}

\maketitle
 
Ptychography is a family of quantitative phase imaging methods which has found applications across a wide range of wavelengths, from visible light \cite{zheng2013,wang2023} through X-ray imaging \cite{thibault2008,pfeiffer2018} and increasingly, electron microscopy \cite{jiang2018a,chen2021}. It's popularity stems from the reliability of phase retrieval from ptychographic data, which can produce quantitative phase images of excellent quality even in cases where high quality lenses are not available.

Single-shot ptychography is a variant of ptychography which was first discussed by Pan et al. in 2013 \cite{pan2013} and later fully developed by Sidorenko et al. in 2016 \cite{sidorenko2016}. In this method, a grid of overlapping light beams illuminates a sample. Each beam approaches the sample from a different angle, allowing the beams to separate and avoid overlapping at the detector plane. In this way, a full ptychographic dataset is captured in a single exposure. The reconstructions can be very reliable due to the robustness of ptychography, even though only one intensity pattern is recorded.

In the years since the initial development at optical wavelengths, several extensions on the core idea have been implemented. Elements such as time-resolved multiplexing \cite{sidorenko2017,veler2024}, polarization resolved multiplexing \cite{chen2018a}, and Fourier ptychography \cite{he2018c} have been added. In addition, interest has developed in implementing single-shot ptychography at X-ray wavelengths. This is driven by the possibility of nanometer-scale resolution and the relative dearth of stable, single-shot quantitative X-ray phase imaging methods for extended samples.

However, the limitations of X-ray optics make it difficult to generate a grid of identical beams with sufficient angular separation and uniform intensities. Despite these challenges, the first single-shot X-ray ptychography reconstructions were reported in 2022 by Kharitonov et al. \cite{kharitonov2022} using a $4 \times 4$ grid of diffraction patterns.

In all of the single-shot ptychography experiments we have discussed so far, the same general strategy was used to solve the phase retrieval problem. The single recorded intensity pattern was first divided into a collection of individual, smaller diffraction patterns, each centered on an individual beam and tagged with a corresponding translation at the sample plane. This collection of diffraction patterns was then reconstructed with an appropriate algorithm for the relevant variant of ptychography.

The main advantage of this approach is that it recovers both the probe and the object, so the probe need not be known a priori. However, the explicit division of the detector into sub-regions also limits the resolution of the final reconstruction by limiting the numerical aperture of the analyzed diffraction data.

It is understood that this is not a fundamental limitation of the information content of the raw data. It is simply imposed by the treatment of that data, which explicitly ignores the possibility of the object having structures at sufficiently small length scales to scatter light between different beams. In cases where the object does contain structure at high frequencies, the scattering from neighboring beams does overlap and interfere, causing reconstructions based on this strategy to perform poorly.

In fact, two recent publications have used machine learning to exceed this resolution limit \cite{wengrowicz2020,wengrowicz2024}. These methods share two additional traits. First, their input is the full detector image. This decouples the final pixel size of the reconstruction from the inter-beam spacing on the detector. Second, they leverage a priori information about the illumination structure, either explicitly \cite{wengrowicz2024} or by encoding it into a trained neural network \cite{wengrowicz2020}.

Neither of these features are fundamentally limited to the deep learning context. In this work we demonstrate how a purely iterative algorithm for single-shot ptychography which uses a pre-calibrated probe and operates on full diffraction patterns, without partitioning them into a ptychography dataset, can also overcome these limitations.

\begin{figure}[ht]
\centering
\fbox{\includegraphics[width=0.97\linewidth]{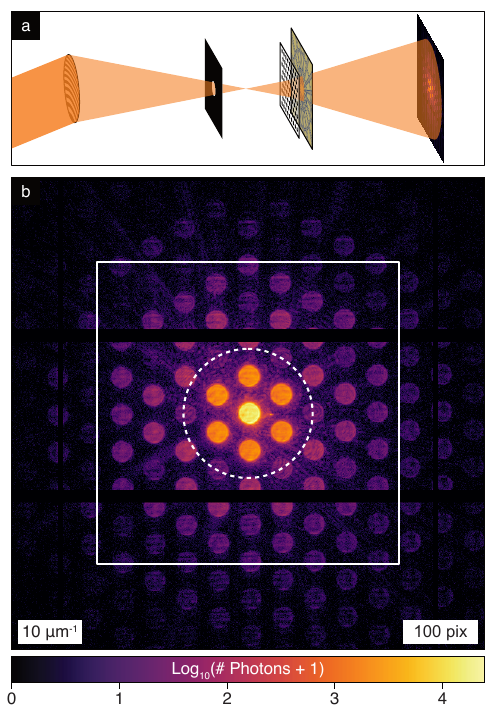}}
\caption{(a) A diagram of the experiment. From left to right: off-axis zone plate, order selecting aperture, diffraction grating, test sample, and detector. (b) The diffraction pattern used for single exposure reconstructions. The solid white rectangle denotes the applied band-limiting constraint. The dashed white circle denotes the final resolution.}
\label{fig:expoverview}
\end{figure}

We demonstrated this method using single-shot X-ray ptychographic data collected at the cSAXS beamline of the Swiss Light Source. In the experiment, diagrammed in Fig. \ref{fig:expoverview}a, a beam of monochromatic $\SI{6.2}{\kilo\eV}$ light was produced with a Si \{111\} double-crystal monochromator and passed through slit with a horizontal aperture of $\SI{100}{\micro\meter}$. $\SI{22}{\meter}$ downstream of the slits, this light passed through a $\SI{50}{\micro\meter}$ diameter pinhole and illuminated an off-axis region of an X-ray zone plate with a $\SI{70}{\nano\meter}$ outer zone width and a nominal focal length of $\SI{70}{\milli\meter}$ at $\SI{6.2}{\kilo\eV}$. The focal spot of this optic was placed $\SI{4.5}{\milli\meter}$ upstream of a hexagonal diffraction grating with a pitch of $\SI{200}{\micro\meter}$. The grating split this light into a grid of beams which propagated a further $\SI{0.955}{\milli\meter}$ downstream before interacting with the sample.

Each individual light beam is identifiable on the raw detector image shown in Fig. \ref{fig:expoverview}b, captured on a Pilatus 2M detector \cite{henrich2009} with $\SI{172}{\micro\meter}$ pixels placed $\SI{7.336}{\meter}$ downstream of the sample. This diffraction pattern was used for all single-exposure reconstructions presented here. From the perspective of single-shot ptychography, the array of spots forms a ptychographic dataset with a $\SI{3.5}{\micro\meter}$ diameter beam scanned over a sample in a hexagonal pattern with an effective step size of $\SI{1.2}{\micro\meter}$.

The data has two features which are devastating for traditional algorithms, but are nevertheless difficult to avoid when working with X-ray optics. First, the intensity of the beams drops by more than an order of magnitude between the zeroth, first, and second orders of diffracted beams. Second, diffraction from the central beam interferes with the outer beams.

\begin{figure}[ht]
\centering
\fbox{\includegraphics[width=0.97\linewidth]{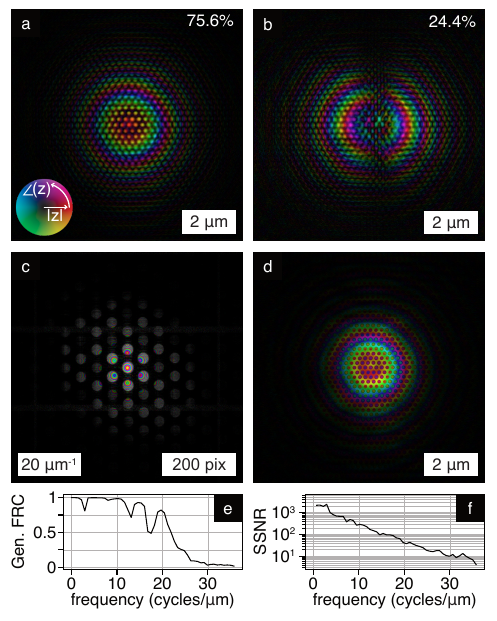}}
\caption{(a,b) The two modes of the recovered probe, plotted in the hsv color space with phase as hue and amplitude as value. The fraction of the total intensity contained in each mode is marked at the top right corner. (c) The Fourier space representation of the top probe mode, with phase plotted as hue and the logarithm of intensity plus one, in units of photons, plotted as value.  (d) The dominant probe mode, propagated to the plane of the grating. (e) A generalized Fourier ring correlation curve \cite{levitan2022} from the recovered probe. (f) The spectral signal-to-noise ratio of the reconstructed object.}
\label{fig:probe_breakdown}
\end{figure}

Instead of attempting to reconstruct this single diffraction pattern as a ptychographic scan without any further information, we start by collecting a full ptychographic dataset as the sample is physically scanned through the illumination. This dataset was analyzed with an automatic differentiation based ptychography algorithm, using a mixed-mode model \cite{thibault2013} with two probe modes. The retrieved probe modes, shown in Fig. \ref{fig:probe_breakdown}a and \ref{fig:probe_breakdown}b, capture the details of the illumination at the sample plane. The fine-scale hexagonal lattice visible within the probe is caused by the coherent superposition of the partially overlapping beams created by the diffraction grating. Viewing the recovered probe in Fourier space, as in Fig. \ref{fig:probe_breakdown}c, highlights the fact that the retrieved illumination function contains information about the full grid of beams, including their relative phase factors.

Propagating the illumination back to the plane of the diffraction grating, as shown in Fig. \ref{fig:probe_breakdown}d, reveals the sharp features of the lithographically patterned grating. A consistency analysis of the recovered probes from two repeat datasets using the generalized Fourier ring correlation \cite{levitan2022}, shown in Fig. \ref{fig:probe_breakdown}e, reveals that the probe is consistently recovered to a half-pitch resolution on the order of $\SI{25}{\nano\meter}$. The object itself was recovered with a sufficiently high signal-to-noise ratio to be used as a ground truth to assess the quality of subsequent single-frame reconstructions,  demonstrated by the spectral signal-to-noise ratio curve plotted in Fig. \ref{fig:probe_breakdown}f.

For comparison to our later results, we next consider what level of quality is achievable with a traditional reconstruction method. To this end, we manually processed the single diffraction pattern into a ptychographic dataset. All beams within the fourth diffraction order of the grating were included, excluding those which were obscured by a detector gap. This makes 43 diffraction patterns in total, $49\times49$ pixels each. We performed a single-shot ptychography reconstruction on this dataset, using an algorithm which corrects for nonuniform illumination intensity, positioning errors at the sample plane, and a shift of each diffraction pattern at the detector plane. The single-shot ptychography reconstruction was additionally stabilized by applying a phase-only constraint to the object.

\begin{figure}[ht]
\centering
\fbox{\includegraphics[width=0.97\linewidth]{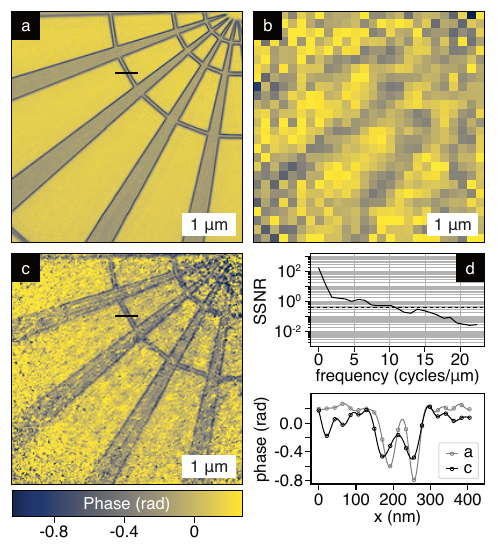}}
\caption{(a) The object recovered via scanning ptychography, downsampled to match the pixel size of the single-frame reconstruction performed with the randomized probe imaging algorithm. (b) A single-frame reconstruction performed using a standard style single-shot ptychography algorithm. (c) A single-frame reconstruction using the randomized probe imaging algorithm. (d) Upper, the spectral signal-to-noise ratio of the image in c, estimated via the Fourier cross resolution \cite{penczek2010}. Lower, line cuts taken from the black lines marked in a and c.}
\label{fig:ssp_recs}
\end{figure}

This reconstruction is shown in Fig. \ref{fig:ssp_recs}b. The object recovered from the calibration scan is shown in Fig. \ref{fig:ssp_recs}a for comparison. The largest-scale features are indeed recovered, but the pixel size of $\SI{174}{\nano\meter}$ is a major limitation and the unaccounted-for inter-beam cross-talk has further degraded the image.

The reconstruction shown in Fig. \ref{fig:ssp_recs}c, performed with the randomized probe imaging algorithm \cite{levitan2020} and the pre-calibrated probe, is a clear improvement. This automatic differentiation based algorithm, defined in \cite{levitan2020}, operates by imposing a band-limiting constraint on the object. In this case, it is limited to the central $400 \times 400$ pixel region of Fourier space indicated in Fig. \ref{fig:expoverview}b. Equivalently, the pixel size of the object was set to $\SI{21.3}{\nano\meter}$. The reconstruction was also stabilized with a phase-only constraint. The forward model for this algorithm is formally written as:

\begin{equation}
\boldsymbol{I} = \sum_{n=1}^N\left| \mathcal{F} \boldsymbol{P}_n \mathcal{F}^{-1} \mathcal{U} \mathcal{F} \exp(i\boldsymbol{T}) \right|^2,
\end{equation}

where $\mathcal{F}$ and $\mathcal{F}^{-1}$ represent the two-dimensional discrete Fourier transform operator and its inverse, respectively. $\boldsymbol{P}_n$ is the discrete representation of the $n$th mode of the pre-calibrated probe. $\mathcal{U}$ is a zero-padding operator, and $\boldsymbol{T}$ is a low-resolution representation of the object's transmission function. This zero-padding operation applies a band-limiting constraint which stabilizes the inverse problem, as discussed in \cite{levitan2020}. To apply an additional phase-only constraint on the object, $\boldsymbol{T}$ is constrained to be purely real. The phase-only constraint can be removed by allowing $\boldsymbol{T}$ to be complex-valued, and the final object function is defined as $\boldsymbol{O} = \exp(i\boldsymbol{T})$.

The spectral signal to noise ratio of the recovered object, estimated via a Fourier cross resolution curve \cite{penczek2010}, is plotted in Fig. \ref{fig:ssp_recs}d. It suggests that a half-pitch resolution of $\SI{50}{\nano\meter}$ was achieved over the roughly $\SI{2.5}{\micro\meter}$ diameter field of view. This is supported by a direct line cut, which clearly distinguishes two iridium-plated edges of a spoke with a separation of less than $\SI{100}{\nano\meter}$.

This is by far the highest resolution achieved to date with single-shot ptychography in any spectral range, and the resolution is 4 times finer than the $\SI{200}{\nano\meter}$ pitch of the diffraction grating used. This is also 3.5 times finer than the numerical aperture limited pixel size of the single-shot ptychography reconstruction. The dashed white line in Fig. \ref{fig:expoverview}b shows the successfully recovered region of the object's Fourier transform. This highlights the fact that, because the probe calibration accurately captures the full structure of the illumination, the inter-beam cross-talk becomes a source of signal rather than a source of noise. 

Our results call attention to the importance of accounting for this cross-talk when working with X-ray data, where the beam intensities span several orders of magnitude. They also serve as a reminder that single-shot ptychography is fundamentally a structured illumination technique, especially when the beams have nontrivial mutual coherence. 

Finally, this methodology also removes a few time-consuming pre-processing steps from the analysis pipeline: the determination of the beam overlaps at the sample position, the calibration of each beam's intensity, and the segmentation of the detector into images corresponding to each beam \cite{kharitonov2022}. All these steps get folded into the ptychography-based calibration of the probe. Consequently, this approach can often be simpler to implement than the traditional single-shot ptychography pipeline.

In summary, we have demonstrated that the reconstruction algorithm used for randomized probe imaging is also applicable to single-shot X-ray ptychographic data, solidifying the previously-discussed connection between the two methods \cite{levitan2020}. It accounts for inter-beam cross-talk, enabling a final half-pitch resolution 3.5 times smaller than the pixel size of a traditional reconstruction. It also removes the limitation that the structure of each beam is identical to all the others. Because of these advantages, the analysis method presented here is likely to perform better than the standard analysis method in most experimental settings, with a particularly large advantage in the X-ray regime.

More generally, the applicability of this simple algorithm to single-shot ptychographic data raises questions about how important it is to preserve the structure of the illumination as a grid of beams. Our results suggest that the key feature of single-shot ptychographic data which enables robust reconstructions is the structure in the probe at high frequencies - not the relationship of the diffraction data to a ptychographic scan. For example, replacing the diffraction grating with an X-ray diffuser with structures on a similar length scale would generate a similar spatial frequency spread in the input illumination, but produce a more even intensity distribution at the sample. However the field evolves, we hope that our results will improve the quality of reconstructions from future single-shot X-ray ptychographic data, while pointing the way to improved illumination strategies.

\begin{backmatter}
\bmsection{Funding}
A.L. acknowledges funding from the European Union’s Horizon 2020 research and innovation programme under the Marie Skłodowska-Curie grant agreement No 884104 (PSI-FELLOW-III-3i). K.W. acknowledges funding from the Swiss National Science Foundation (SNF), grant 166304. Z.G. acknowledges funding received from the SNF, project number 200021\_178788. O.C. acknowledges funding received from the European Research Council (ERC), grant 819440-TIMP.

\bmsection{Acknowledgments}
Many thanks to Wenhui Xu, who gave insightful comments on an early version of this paper.

\bmsection{Disclosures}
The authors declare no conflicts of interest.

\bmsection{Data availability}
Data and analysis code underlying the results presented in this paper are available in Ref. \cite{levitan2024a}.

\end{backmatter}

\bibliography{bibliography}

% Full bibliography added automatically for Optics Letters submissions; the following line will simply be ignored if submitting to other journals.
% Note that this extra page will not count against page length
\bibliographyfullrefs{bibliography}

\end{document}